\documentclass[10pt,conference]{IEEEtran}
\IEEEoverridecommandlockouts
\usepackage{orcidlink}
\usepackage{hyperref}
\usepackage{cite}
\usepackage{amsmath,amssymb,amsfonts}
\usepackage{algorithmic}
\usepackage{graphicx}
\usepackage{textcomp}
\usepackage{xcolor}
\usepackage{threeparttable}
\usepackage{multirow}
\usepackage{booktabs}
\usepackage{makecell}
\usepackage{enumitem}
\usepackage{balance}
\usepackage{tcolorbox}
\tcbuselibrary{skins}

\makeatletter
\newcommand{\linebreakand}{%
  \end{@IEEEauthorhalign}
  \hfill\mbox{}\par
  \mbox{}\hfill\begin{@IEEEauthorhalign}
}
\makeatother

\newcommand\add[1]{\textcolor{black}{#1}}

\author{
\IEEEauthorblockN{
Taiming Wang\IEEEauthorrefmark{2}\,\orcidlink{0000-0001-7350-4419}, 
Yanjie Jiang\IEEEauthorrefmark{3}\IEEEauthorrefmark{1}\thanks{* Corresponding author: Yanjie Jiang (yanjiejiang@tju.edu.cn)}\,\orcidlink{0000-0001-6404-9143}, 
Chunhao Dong\IEEEauthorrefmark{2}\,\orcidlink{0000-0002-7560-2685}, 
Yuxia Zhang\IEEEauthorrefmark{2}\,\orcidlink{0000-0002-9371-5931}, and 
Hui Liu\IEEEauthorrefmark{2}\,\orcidlink{0000-0002-3267-6801}
}
\IEEEauthorblockA{
\IEEEauthorrefmark{2}School of Computer Science and Technology, Beijing Institute of Technology, Beijing, 100081, China\\
\IEEEauthorrefmark{3} College of Intelligence and Computing, Tianjin University, Tianjin, 300072, China \\
Email:\{wangtaiming, dongchunhao22, yuxiazh, liuhui08\}@bit.edu.cn, yanjiejiang@tju.edu.cn
}
}

\begin{document}

\title{Wired for Reuse: Automating Context-Aware Code Adaptation in IDEs via LLM-Based Agent\\}


\maketitle

\begin{abstract}
\textit{Copy-paste-modify} is a widespread and pragmatic practice in software development, where developers adapt reused code snippets, sourced from platforms such as Stack Overflow, GitHub, or LLM outputs, into their local codebase. A critical yet underexplored aspect of this adaptation is \textit{code wiring}\add{ : the context-aware process of substituting unresolved variables in pasted code with suitable variables or expressions from the surrounding context.} Existing solutions either rely on heuristic rules or historical templates, often failing to effectively utilize contextual information, despite studies showing that over half of adaptation cases are context-dependent.
In this paper, we introduce \emph{WIRL}, an LLM-based agent for code wiring framed as a Retrieval-Augmented Generation (RAG) infilling task. \emph{WIRL} combines an LLM, a customized toolkit, and an orchestration module to identify unresolved variables, retrieve context, and perform context-aware substitutions. To balance efficiency and autonomy, the agent adopts a mixed strategy: deterministic rule-based steps for common patterns, and a state-machine-guided decision process for intelligent exploration.
We evaluate \emph{WIRL} on a carefully curated, high-quality dataset consisting of real-world code adaptation scenarios. Our approach achieves an exact match precision of 91.7\% and a recall of 90.0\%, outperforming advanced LLMs by 22.6 and 13.7 percentage points in precision and recall, respectively, and surpassing IntelliJ IDEA by 54.3 and 49.9 percentage points. These results underscore its practical utility, particularly in contexts with complex variable dependencies or multiple unresolved variables. We believe \emph{WIRL} paves the way for more intelligent and context-aware developer assistance in modern IDEs.
\end{abstract}


\section{Introduction}\label{sec:Introduction}
Copy-paste-modify is a common and inevitable practice during development. The developers frequently reuse code snippets from online programming Q\&A communities, e.g., Stack Overflow~\cite{StackOverflow}, open-source repositories, e.g., GitHub~\cite{GitHub}, or code generation of LLMs. However, in most cases the reused code snippets need additional adaptations to integrate to the local codebase. As reported, more than 85\% code snippets from Stack Overflow need to be adapted (modified) before integration into the local code~\cite{Zhang2024How,Chen2024How}.   Variable identifiers are usually under adaptation since undeclared or conflict identifiers would lead to compilation errors or potential bugs. Consequently, code wiring, replacing unresolved variables with existing ones from the local context, is one of the most prevalent forms of adaptation. An example of code wiring practice is presented in Fig.~\ref{fig:MotivatingExample}. 
Automatic code wiring can help developers get rid of the repetitive and error-prone processes during the adaptation and concentrate on the complex business logic. 
However, existing approaches are designed based on either simple heuristic rules~\cite{IntelliJIDEA} or templates extracted from historical modifications~\cite{ExampleStack}, leaving the context not effectively leveraged although 56.1\% of the adaptation cases are dependent on the surrounding context as reported by Zhang et al.~\cite{Zhang2024How}. \add{Advanced code editing approaches~\cite{GrACE,CoEdPilot} are dependent on a rich history of prior edits, information that is inherently scarce in the context of code wiring.} Full-parameter LLMs have demonstrated promising performance; however, their high latency renders them impractical for real-world applications. On the other hand, distilled LLMs with smaller parameter sizes tend to under perform in terms of accuracy and robustness. 

To this end, we present \emph{WIRL}, an LLM-based agent for code \textbf{W}iring through \textbf{I}nfilling with \textbf{R}AG and \textbf{L}LMs. \emph{WIRL} consists of three core components: an LLM, a customized toolkit, and an agent pilot. The customized toolkit offers three primary functionalities: (1) identifying and locating unresolved elements, (2) analyzing and collecting contextual information, and (3) infilling and recommending suitable substitutions. The agent pilot is responsible not only for initializing the prompt but also for coordinating the interaction between the LLM agent and the toolkit. It parses the LLM's output, updates the prompt dynamically, and invokes the appropriate tools as needed.
With a dynamically updated prompt, \emph{WIRL} incrementally gathers relevant context by invoking appropriate tools until the agent determines that sufficient information has been collected to make a final recommendation. To fully leverage the capabilities of LLMs, we reformulate the adaptation task as a retrieval-augmented generation (RAG)-based infilling task for the unresolved elements, framing it as a code completion problem which aligns more naturally with the strengths of LLMs. Empirical evidence from Zhang et al.~\cite{zhang2024instructinteractexploringeliciting} supports this reformulation, showing that LLM performance on code snippet adaptation tasks lags behind their performance on code completion and generation tasks.
Moreover, by equipping with the RAG-based infilling strategy, even distilled LLMs with smaller parameter sizes can outperform full-parameter LLMs, thereby meeting both the latency and accuracy requirements of real-world software development scenarios.
To enhance the efficiency of \emph{WIRL}, we adopt a hybrid execution mode for the agent, wherein essential steps are extracted and executed initially without invoking the LLM. In addition to this optimization, we introduce a state machine to guide the agent’s exploration process, thereby reducing unnecessary or ineffective attempts and improving overall execution efficiency.

To assess the efficiency of \emph{WIRL}, we manually curated a high-quality code wiring dataset derived from a publicly available code adaptation dataset~\cite{CWDataset}. Evaluated on this dataset, \emph{WIRL} achieves an exact match precision of 91.7\% and a recall of 90.0\%. Notably, it outperforms advanced LLM baselines by 22.6 and 13.7 percentage points in precision and recall, respectively, and exceeds the performance of IntelliJ IDEA by 54.3 and 49.9 percentage points. These results highlight the effectiveness and practical advantage of \emph{WIRL}. We also analyzed time efficiency and token consumption, demonstrating that \emph{WIRL} meets real-time requirements with affordable computational cost.

The contributions of this paper are as follows:
\begin{itemize}
\item We propose \emph{WIRL}, an LLM-based agent specifically designed for context-aware code wiring recommendations.
\item We develop a customized toolkit that provides essential functionalities for precise and efficient code adaptation.
\item We construct and release a high-quality dataset for code wiring evaluation, curated and validated through careful manual inspection.
\end{itemize}

The rest of this paper is structured as follows. Section~\ref{sec:MotivatingExample} motivates this study. Section~\ref{sec:Approach} illustrates the design of \emph{WIRL}. Section~\ref{sec:Evaluation} presents experimental details and results. \add{Section~\ref{sec:Discussion} first explores the broader implications and robustness of our findings through analysis of applicability and versatility. It then discusses potential threats to validity and the inherent limitations of our study.} Section~\ref{sec:RelatedWork} discusses the related works, and Section~\ref{sec:Conclusion} concludes this paper.

\section{Motivating Example}\label{sec:MotivatingExample}
\begin{figure}
    \centering
\includegraphics[width=\linewidth,clip]{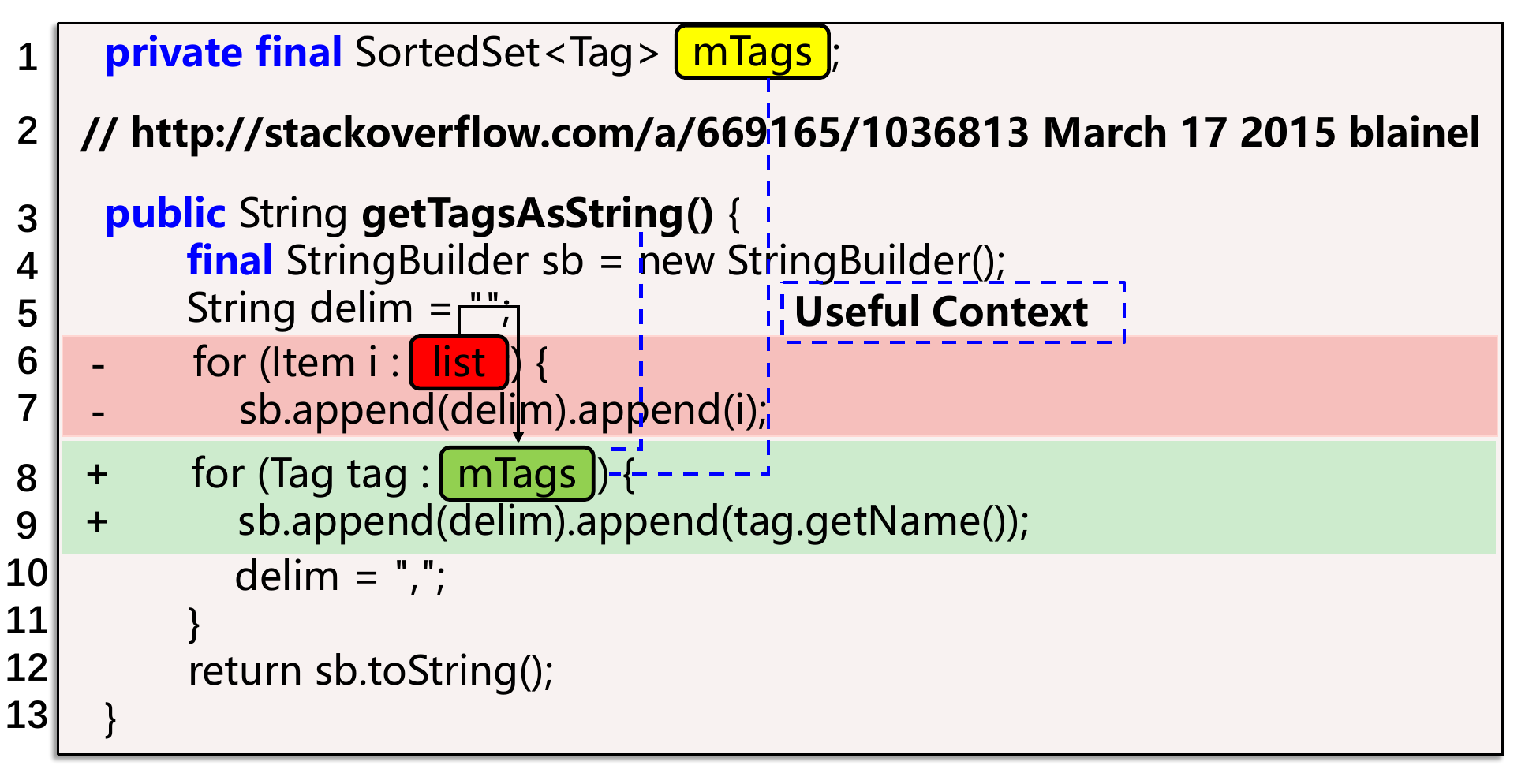}
    \caption{Motivating Example}
    \label{fig:MotivatingExample}
\end{figure}
To provide the intuition about how \emph{WIRL} works and to define the key terminologies, we begin with a motivating example of a code wiring instance.
Fig.~\ref{fig:MotivatingExample} presents an example of copy-paste-modify (specifically code wiring) practice where a developer reused a piece of code snippet from Stack Overflow~\cite{StackoverFlowURL} and adapted it to the GitHub local repository~\cite{GitHubURL}. The developer copied and pasted the line 6 and line 7 and then modified them into line 8 and line 9. For the sake of clarity, in this paper we call the code snippets in line 6 and line 7 \emph{\textbf{isolated code}} since it is isolated before integration. The code snippets in line 8 and line 9 are called \emph{\textbf{integrated code}}, i.e., code after the integration. Specifically, in this example, local variable \emph{``list''} (highlighted in red) is undefined and leads to syntax errors. The developers substituted it with the predefined field variable \emph{``mTags''}(highlighted in yellow and green) to resolve the syntax errors. We call \emph{``list''} an \emph{\textbf{unresolved element}}. It is worth noting that variables are not the only code entity that need to be modified, literal values or a method invocation expression could also be the target code entity. Consequently, we use \emph{``element''} instead of only \emph{``variable''}.
\emph{``mTags''} is called the \emph{\textbf{context element}}, i.e., the code elements in the context. Similarly, a \emph{context element} could also be either a variable (including local variables, parameters, and fields) or an expression (e.g., method invocation).

We first present the problem formulation, which serves as an important foundation of our approach.
Assume the set of \emph{unresolved elements} is denoted by $\mathcal{U} = {u_i}$ and the set of \emph{context elements} by $\mathcal{C} = {c_i}$. The code wiring task can be formally defined as identifying an injective mapping $\mathcal{M}: \mathcal{U} \rightarrow \mathcal{C}$, where each $u_i \in \mathcal{U}$ is mapped to a distinct $c_i \in \mathcal{C}$. In this work, however, we reformulate code wiring as a retrieval-augmented generation (RAG)-based infilling task. Specifically, each $u_i \in \mathcal{U}$ is replaced with a placeholder in the \emph{isolated code}. The objective then becomes to infill these placeholders using the retrieved context information, including the variable names themselves, which is motivated by the observation that literal similarity often provides valuable contextual cues.
This reformulation is grounded in the insight that code completion is more naturally aligned with the capabilities of LLMs than explicit code adaptation. This design choice is further supported by recent findings by Zhang et al.~\cite{zhang2024instructinteractexploringeliciting}, which highlight the superior performance of LLMs in code completion tasks relative to direct adaptation scenarios.

In this example, \emph{WIRL} first identifies the unresolved variable \emph{``list''}, and then proceeds to iteratively collect relevant contextual information such as available variables in the current scope, semantic hints indicating that \emph{``list''} should refer to a collection object, and method names suggesting that \emph{``Tags''} is the intended target. Then it assesses the suitability of candidate variables by invoking a set of external tools tailored for context analysis. Once the context is deemed sufficient, \emph{WIRL} conduct infilling and \emph{``mTags''} is filled as the appropriate variable here. Eventually, \emph{WIRL} automatically applies the modification within the IDE for developers.

\begin{figure*}
    \centering
\includegraphics[width=\linewidth,clip]{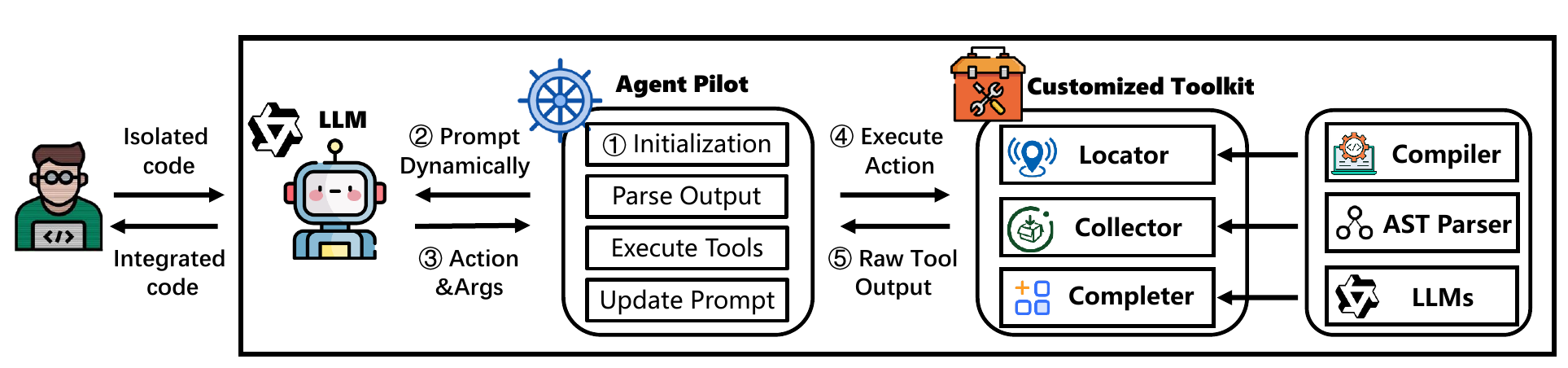}
    \caption{Overview of WIRL}
    \label{fig:overview}
\end{figure*}

\section{Approach}\label{sec:Approach}

\subsection{Overview}
The overview of \emph{WIRL} is illustrated in Fig.~\ref{fig:overview}. \emph{WIRL} consists of three core components: an \emph{LLM}, a \emph{customized toolkit} containing three types of tools, i.e., \emph{locator}, \emph{collector}, and \emph{completer}, and an \emph{agent pilot}, which orchestrates communication between the LLM and the toolkit by parsing outputs, updating prompts, and executing tool invocations.
Given a piece of \emph{isolated code}, the \emph{agent pilot} begins by initializing the prompt (\emph{Step 1}). This prompt includes system roles, task descriptions, and instructions on how the task should be addressed using the available tools. It also embeds the location of \emph{unresolved elements} and results from preliminary analyses.
Using this prompt, the \emph{agent pilot} issues a request to the LLM (\emph{Step 2}). The LLM responds by specifying a tool to invoke from the customized toolkit (\emph{Step 3}). The \emph{agent pilot} then parses the response and executes the corresponding tool with the appropriate arguments (\emph{Step 4}). The tool's output is subsequently parsed and integrated into the prompt (\emph{Step 5}) for the next iteration.
This loop continues iteratively until either the \emph{completer} tool is executed, or a predefined computational budget is reached. \add{The customized toolkit, implemented by the proposed approach,  depends on high-level abstractions like compilers, AST parsers, and LLMs. }

\subsection{Dynamic 
Prompting}~\label{subsec:DynamicPrompting}
Adopting the \emph{ReAct} framework~\cite{Yao2023ReAct}, \emph{WIRL} iteratively reasons and acts until the goal is met or the budget is exceeded. In each cycle, the \emph{agent pilot} parses tool outputs and integrates them into the prompt for the next iteration, which consists of both static and dynamic sections.

\subsubsection{Role (Static)}
The static section defines the agent's role as a Senior Java Development Engineer, tasked with autonomous, variable-level code integration based on the provided context, explicitly prohibiting any user intervention.
\subsubsection{Goals (Static)}
Three primary goals to accomplish:
\begin{itemize}[leftmargin=1em]
\item \textbf{\emph{Analyze context:}} Evaluate the current code context to determine whether it provides sufficient information to suggest a valid substitution.
\item \textbf{\emph{Collect more context:}} If the existing context is inadequate, invoke the appropriate tools from the provided toolkit to gather additional relevant information.
\item \textbf{\emph{Suggest a substitution:}} Once adequate context is obtained, leverage the agent’s completion tool to recommend a suitable substitution for the unresolved element.
\end{itemize}

\subsubsection{Guidelines (Static)}
Guidelines the agent should adhere:
\begin{itemize}[leftmargin=1em]
    \item \emph{Limited budget:} We emphasize that the budget is limited and it is important to be efficient to select the next steps.
    \item \emph{Type should match:} We inform the agent that the basic principle is to make sure the retrieved \emph{context element} is type-compatible with \emph{unresolved element}.
    \item \emph{Be careful to stretch:} More context typically leads to higher latency. Consequently, the agent is instructed to complete its recommendation using the minimal necessary context and to extend the context scope cautiously.
    \item \emph{Extend scope:} We instruct the agent that in some cases, no suitable substitution candidates may be found within the immediate context. In such scenarios, the agent should expand its search scope beyond the current class, as the \emph{context element} is more likely to be a method invocation expression rather than a simple variable.
\end{itemize}
\subsubsection{State Description (Dynamic)}
To guide the agent in utilizing the \emph{customized toolkit} in a meaningful and efficient manner, we define a state machine comprising three states: \emph{initial state}, \emph{insufficient context state}, and \emph{sufficient context state}. Each state is associated with a set of available tools, as described in Section~\ref{subsec:ToolDescription}, and the state description section of the prompt informs the agent of its current status. \add{This mechanism is introduced because, in our preliminary experiments, we observed that the agent often engaged in aimless exploration when operating without explicit guidance. 
The transition rules between each state are presented in Table~\ref{tab:TransitionRules}. 
The agent begins in the \emph{initial state}, where it identifies unresolved elements and performs a preliminary analysis of the context.  It then transitions to the \emph{insufficient context state}  to iteratively gather necessary contextual information. The agent moves to the \emph{sufficient context state}  once enough information has been gathered or a predefined resource budget (i.e., iteration limit) is exhausted. In this state, it invokes the completer to generate the recommendation before terminating (\emph{done state}). This structured workflow, with its built-in budget to handle errors and prevent infinite loops, ensures a focused and efficient decision-making process.}
It is important to note that while the state machine offers structured guidance, it does not enforce a fixed sequence of tool invocations. The selection and ordering of tools are left entirely to the agent’s discretion.
\begin{table}[]
\centering
\caption{Transition Rules of State Machine}
\begin{tabular}{@{}lll@{}}
\toprule
From State           & Transition Rules                                                                         & To State             \\ \midrule
Initial              & Initialization Finished                                                                  & Insufficient Context \\
Insufficient Context & \begin{tabular}[c]{@{}l@{}}Budget Exhausted or \\ Context Deemed Sufficient\end{tabular} & Sufficient Context   \\
Sufficient Context   & Completer Executed                                                                       & Done               \\ \bottomrule
\end{tabular}
\label{tab:TransitionRules}
\end{table}
\subsubsection{Available Tools (Dynamic)} 
This section illustrates the available tools (refer Section~\ref{subsec:ToolDescription}) agent can call in each state.

\subsubsection{Gathered Information (Dynamic)}
A key capability of \emph{WIRL} lies in its ability to gather contextual information through tool execution, which forms the foundation for generating accurate recommendations. To ensure the agent remains aware of previously executed tools and the context collected, this section of the prompt records the reasoning thoughts, the name of tool, its corresponding arguments (parsed from the LLM’s output), along with the results returned by tool invocations (parsed from the tools' output).

\subsubsection{Output Format (Static)}
To enable the \emph{agent pilot} to effectively parse the output, we require all responses from the LLM be structured in JSON format. Each JSON object must contain three mandatory fields: \emph{``thought''}, \emph{``action''}, and \emph{``action\_input''}. The \emph{``thought''} field captures the agent’s reasoning and decision-making process based on the current context. The \emph{``action''} field specifies the tools' name to be invoked next, while the \emph{``action\_input''} field provides necessary arguments for that tool, also formatted as a JSON object.

\subsection{Customized Toolkit}\label{subsec:ToolDescription}
\begin{table*}[]
\centering
\caption{ Available Tools for the Agent}
\begin{tabular}{@{}llll@{}}
\toprule
Tools           & Type      & Applicable State               & Description \\ \midrule
identify\_unresolved\_elements  & Locator  & Initialization      & Analyze compiler information to identify the \emph{unresolved elements}.\\ 
get\_available\_variables   & Collector  & Initialization      &  Conduct code analysis and get the available variables in the current context scope.       \\
get\_unused\_variables   & Collector    & Initialization       &  Conduct data flow analysis and get the unused variables in the current context.            \\
is\_argument         & Collector      & Initialization         &  Judge whether the \emph{unresolved element} plays an argument in a method invocation.           \\
is\_receiver         & Collector       & Initialization        & Judge whether the \emph{unresolved element} plays a receiver in a method invocation.  \\          
retrieve\_identical\_function\_call &Collector & Insufficient Context&  Retrieve variables that invoke the identical function calls with \emph{unresolved element}.           \\
reserve\_type\_compatible\_ones &Collector & Insufficient Context    &  Only reserve the variables that are type-compatible as the \emph{unresolved element}.          \\
sort\_by\_literal\_similarity  &Collector & Insufficient Context    &  Sort the available variables by their similarity with the \emph{unresolved element}.         \\
get\_method\_names   &Collector   & Insufficient Context            &  Collect method members with same type as \emph{unresolved element} in the given class.           \\
execute\_completion      & Completer     & Sufficient Context     &  Invoke LLM to complete the code snippets with the collected context information.           \\ \bottomrule
\end{tabular}
\label{tab:ToolDescription}
\end{table*}
One of the key novelties of \emph{WIRL} is its ability to autonomously decide which tools to invoke based on the current state. The toolkit provided for \emph{WIRL} (as shown in Table~\ref{tab:ToolDescription}) is carefully customized to facilitate the code wiring task.

\subsubsection{Identifying unresolved elements} A prerequisite for effective \emph{code wiring} is the accurate identification of \emph{unresolved elements}. The \emph{locator} tool (\emph{identify\_unresolved\_elements}) leverages compiler information to detect unresolved variables and their references, and it is executed as the initial step. Since identifying and locating these elements is fundamental to addressing the task, the \emph{agent pilot} invokes the \emph{locator} automatically, without requiring input from the LLM.

\subsubsection{Collecting Context Information} To minimize unnecessary LLM invocations and reduce iteration overhead, we provide two tools, i.e., \emph{get\_available\_variables} and \emph{get\_unused\_variables}, to help the agent efficiently construct a list of candidate variables. The \emph{get\_available\_variables} tool parses the AST of the current Java file and collects all accessible local variables, method parameters, and class fields. To reflect realistic development scenarios, only local variables declared before the \emph{unresolved element} are considered valid candidates.
Once available variables are gathered, the \emph{get\_unused\_variables} tool performs data-flow analysis to examine the usage of each variable. Variables with no detected references are marked as unused.
To determine the syntactic role of an \emph{unresolved element}, the tools \emph{is\_argument} and \emph{is\_receiver} inspect its AST context to identify whether it is enclosed within a method invocation. If the element appears as a method argument, its expected type can be inferred from the corresponding formal parameter, and the parameter name itself provides useful semantic cues. If the element is the receiver of a method call, the tool returns the \emph{invoked method member} to support further analysis.
The \emph{reserve\_type\_compatible\_ones} tool filters candidate variables by retaining only those whose types are compatible with that of the \emph{unresolved element}. When the unresolved element is identified as a method receiver, the \emph{retrieve\_identical\_function\_call} tool takes the \emph{invoked method member} as input and searches the current file for instances of the same member invocation, which may reveal contextually relevant patterns or variable usages.
Finally, if no valid candidates are found in the current scope, the \emph{get\_method\_names} tool is invoked. Given a class name, it returns the method members of that class that match the type of the \emph{unresolved element}, providing additional opportunities for accurate substitution.
The \emph{sort\_by\_literal\_similarity} tool is invoked when the agent determines that lexical similarity is a relevant factor. This tool computes the Levenshtein distance between the \emph{unresolved element} and each candidate variable, ranking the candidates based on their literal similarity.

\subsubsection{Infilling Isolated Code}
Once the agent has collected sufficient contextual information, it invokes the \emph{execution\_completion} tool to infill the placeholders using the gathered data. This tool is implemented via an LLM call, where the prompt corresponds to the final version of the dynamically updated prompt that incorporates all previously collected context information.

\subsection{Agent Pilot}
The \emph{agent pilot} orchestrating the communication between \emph{LLM} and \emph{customized toolkit}, playing a essential role in \emph{WIRL}. Its responsibilities are outlined as follows:

\subsubsection{Initializing Prompt} The \emph{agent pilot} begins by initializing the prompt with static components such as the role definitions and task objectives. It then incorporates the input code (as described in Section~\ref{subsubsec:InputOutputFormat}) into the prompt.
Subsequently, the prompt is dynamically updated with the locations of \emph{unresolved elements} and relevant contextual information, obtained by invoking the \emph{Locator} (\emph{identify\_unresolved\_elements}) and \emph{Collector} tools (\emph{get\_available\_variables}, \emph{get\_unused\_variables}, \emph{is\_argument}, and \emph{is\_receiver}).
This initialization process substantially reduces unnecessary LLM invocations and accelerates context analysis, thereby enhancing the overall efficiency of \emph{WIRL}.

\subsubsection{Parsing Output}
Since the outputs of LLMs are returned in JSON format, they can not be directly used for tool execution. Accordingly, the second responsibility of the \emph{agent pilot} is to validate and process these outputs. If an LLM response deviates from the expected structure, e.g., due to hallucinations or formatting errors, the \emph{agent pilot} must attempt to correct the output or handle exceptions gracefully. In addition to output parsing and error handling, the \emph{agent pilot} also maintains a memory of previously executed tools and their corresponding arguments to prevent redundant operations.

\subsubsection{Executing Tools}
With the parsed tool names and corresponding arguments, the \emph{agent pilot} invokes the designated tools and returns their outputs. Notably, all tool executions are performed within an isolated environment to ensure they do not interfere with the internal functioning of \emph{WIRL}.                 

\subsubsection{Updating Prompt}
Once the tool output is available, the \emph{agent pilot} updates all dynamic sections of the prompt in preparation for the next iteration. Specifically, it modifies the current state and the list of available tools, and appends the \emph{``thought''}, \emph{``action''}, \emph{``action\_input''}, and \emph{``observation''} (i.e., the tool’s output) to the accumulated context information.

\section{Evaluation}\label{sec:Evaluation}
\subsection{Research Questions}
\begin{itemize}[leftmargin=1em]
    \item \textbf{RQ1:} How well does WIRL perform compared against baselines?
    \item \textbf{RQ2:} How well does WIRL perform regarding latency?
    \item \textbf{RQ3:} How well does WIRL perform regarding token consumption and monetary cost?
    \item \add{\textbf{RQ4:} How do the design components contribute to the performance of WIRL?}
\end{itemize}
RQ1 investigates the effectiveness of \emph{WIRL} in comparison to selected baseline approaches for automatic code wiring. By addressing this question, we aim to understand how well \emph{WIRL} performs in real-world development scenarios, as the evaluation leverages real data under realistic experimental conditions. Additionally, this analysis helps identify specific cases where \emph{WIRL} outperforms the baselines. Gaining insights into the strengths of \emph{WIRL} can guide future improvements and optimizations of the approach.

RQ2 examines the time efficiency of \emph{WIRL} relative to baseline methods, with a particular focus on whether \emph{WIRL} can meet the responsiveness requirements of integrated development environments (IDEs) and developers. To answer this question, we use the average time taken to complete a code wiring task as the primary metric. This allows us to assess whether \emph{WIRL} delivers timely recommendations that support fluid and productive software development workflows.

RQ3 assesses the token consumption and associated monetary cost of \emph{WIRL}. Specifically, we evaluate input/output token usage and the total cost incurred per code wiring instance. By comparing these metrics against those of the baseline approaches, we aim to determine the cost-efficiency of \emph{WIRL} in delivering practical code wiring support at a reasonable computational expense.

\add{RQ4 quantifies the contribution of each of \emph{WIRL}'s design components, namely the Locator, Collector, and Completer tools, as well as its architectural design, to the system's overall performance. Understanding the individual impact of these elements is crucial for validating the novelty of our approach and justifying the design of the \emph{WIRL} framework.}

\subsection{Dataset}~\label{subsec:Dataset}
To the best of our knowledge, there are no existing datasets specifically collected for the evaluation of automatic code wiring approaches. 
Recently, Zhang et al.\cite{Zhang2024How} investigated context-based code snippet adaptation and constructed a high-quality dataset through a combination of automated processing and meticulous manual curation. The resulting dataset comprises 3,628 real-world code reuse cases, where code snippets were copied from Stack Overflow and subsequently adapted for use in GitHub projects. Their dataset was initially derived from the latest version of SOTorrent\cite{Baltes2018SOTorrent} (version 2020-12-31), available via Zenodo~\cite{Zenodo}, and includes only those reuse cases in which GitHub files contain explicit references to Stack Overflow posts.
Given the quality and relevance of this dataset, we adopted it as the foundation for our evaluation and performed additional filtering and manual curation. The construction of the evaluation dataset used in this paper involves the following steps:

\begin{itemize}[leftmargin=1em]
    \item To prevent data leakage and potential overfitting, we first excluded the 300 sampled instances used by Zhang et al.~\cite{Zhang2024How}, as the design of \emph{WIRL} was partially inspired by their empirical observations.
    \item From the remaining 3,328 reuse cases, we performed a deduplication step to eliminate redundancy caused by forked repositories. Subsequently, we constructed a mapping between GitHub repositories and the corresponding Stack Overflow posts.
    \item As \emph{WIRL} depends on compiler feedback and AST binding analysis, it requires the project to be resolvable. Therefore, we conducted a project-wise manual inspection of the remaining dataset. The mappings were sorted in descending order based on the frequency of references to each repository, prioritizing more frequently reused code.
    \item Following this order, two authors independently reviewed each code reuse case to determine whether it qualifies as a \emph{code wiring} instance. In cases of disagreement, discussions were held until a consensus was reached. Ultimately, we identified 100 \emph{code wiring} cases involved with 221 pairs of \emph{unresolved elements} and \emph{context elements} from Stack Overflow to GitHub. For ease of reference, we refer to this curated evaluation dataset as $CWEvaluation$.
\end{itemize}

\subsection{Selected Baselines}
\subsubsection{SOTA Approaches}
\add{Since code editing task is potential to solve the code wiring problem, we include two state-of-the-art approaches in code editing, e.g., GrACE~\cite{GrACE} and CoEdPilot~\cite{CoEdPilot}.}

We selected \emph{ExampleStack}~\cite{Zhang2019Analyze} as a baseline because it represents the state-of-the-art in automatic code wiring. \emph{ExampleStack} is a Google Chrome extension designed to assist developers in adapting and integrating online code examples into their own code repositories. The authors provide comprehensive reproduction instructions, which allowed us to successfully set up and run \emph{ExampleStack} in our evaluation.

Besides that, we selected the widely adopted industrial IDE, \emph{IntelliJ IDEA}~\cite{IntelliJIDEA}, as one of our baselines. The renaming recommendation support in IDEA is both practical and suitable to address the code wiring problem.

\subsubsection{Raw LLMs}
We take the following raw LLMs as our baselines:
Distilled models, i.e., \emph{GPT-4o-mini}~\cite{GPT-4o-mini}, \emph{Qwen2.5-Coder-14B}~\cite{Qwen2.5coder14b}, and \emph{Qwen2.5-Coder-32B}~\cite{Qwen2.5coder32b}; Full-parameter models, i.e., \emph{DeepSeek-V3}~\cite{DeepSeekV3} and \emph{Qwen-max}~\cite{QwenMax}.

\subsection{Metrics}~\label{subsec:Metrics}
We adopted the evaluation metrics used by Wang et al.~\cite{Wang2025Recommending} to assess the performance of \emph{WIRL} and the baselines. The definitions are introduced in the following:
\begin{itemize}[leftmargin=1em]
    \item \emph{\#Total Cases:}  the number of cases involved in the evaluation, i.e., number of \emph{unresolved elements}.
    \item \emph{\#Recommendation \add{(\#Rec)}:}  the number of cases where the evaluated approaches make a recommendation for the developers. Note that the number of recommendations (\#recommendation) may not always match the total number of cases (\#total cases). 
    \item  \emph{\#Exact Match \add{(\#EM)}:}  the number of cases where the recommended code elements are identical to the ground truth (i.e., the existing names in context).  
    \item  \emph{$EM_{Precision}$ \add{($EM_{P}$)}:}  the number of exact matches divided by the number of recommendations, i.e., \begin{equation}
         \label{eqt:Precision}
             EM_{Precision}= \frac{\#Exact\ Match}{\#Recommendation}
         \end{equation}
    \item  \emph{$EM_{Recall}$ \add{($EM_{R}$)}:}  the number of exact matches divided by the number of cases involved in the evaluation.
         \begin{equation}
         \label{eqt:recall}
             EM_{Recall}=  \frac{\#Exact\ Match}{\#Total\ Cases}
         \end{equation}
\end{itemize}

\subsection{Experimental Setup}
\subsubsection{Input and Output Format for LLMs}\label{subsubsec:InputOutputFormat}
To better simulate real-world development scenarios, the input code format is configured as follows: 
\begin{itemize}[leftmargin=1em]
    \item  For the LLM baselines, the entire class containing the \emph{isolated code} is provided as input. The \emph{isolated code}, representing the segment requiring adaptation, is explicitly marked using a pair of control tokens, \emph{$<start>$} and \emph{$<end>$}, to delineate the adaptation region. \add{The processed data format and the complete prompt format used for the LLM baselines are available in our public repository~\cite{WIRL}.}
    \item To enhance efficiency, the input to \emph{WIRL} consists solely of the method declaration containing the \emph{isolated code}. \emph{WIRL} is designed to operate with the minimal necessary context and selectively expand the context scope when needed.
    \item Furthermore, we assume that any code following the \emph{isolated code} is unavailable, reflecting the common top-down coding pattern observed during software development. In this scenario, only the code preceding the adaptation region, which within the same method declaration, is accessible as context.
\end{itemize}
For the sake of clarity, we constrained the LLM baselines to output only the \emph{unresolved elements} along with their corresponding \emph{context elements}.

\subsubsection{\add{LLM Configurations}}
To promote self-consistency and mitigate output variability, we set the temperature to 0. \add{The top-p is set to 0.2, and the other parameters are set to default. }Furthermore, each evaluated approach was executed five times, and the most frequently generated output was selected as the final result, following the protocol outlined by Chen et al.~\cite{Chen2023Self}.

\subsubsection{\add{Baseline Configurations}}
\add{As \emph{IDEA}, \emph{ExampleStack}, and \emph{CoEdPilot}} produce ranked lists of candidate substitutions, we only considered their top-ranked recommendation as the final answer to ensure a fair and consistent comparison across all evaluated approaches. \add{For \emph{CoEdPilot}, we also equipped it with our Locator module to maximize its performance.}

\add{Due to the deprecation of the \emph{code-davinci-002} model, which was originally used in the zero-shot implementation of \emph{GrACE}, we sought a suitable replacement. OpenAI officially discontinued support for the Codex API, including \emph{code-davinci-002}, on March 23, 2023, and recommended transitioning to more advanced models. To this end, we opted to use a more advanced model, \emph{GPT-4o}, for our implementation of \emph{GrACE}.}

\subsubsection{Implementations}
To implement \emph{WIRL}, we built a plugin for IntelliJ IDEA using the \emph{IntelliJ PlatForm Plugin SDK}~\cite{IntelliJPluginSDK} and the Langchain4J~\cite{Langchain4J} framework to interface with LLM APIs. 
While \emph{WIRL} is designed to be compatible with any advanced LLMs, in this study we use \emph{Qwen2.5-Coder-14B} as the backend model for two key reasons. First, if \emph{WIRL} demonstrates superior performance over state-of-the-art full-parameter LLMs while relying on a smaller model, it further highlights the strength of our design. Second, models with fewer parameters typically respond faster~\cite{li2025largelanguagemodelinference,Zhu2024ASurvey}, making them more suitable for real-time usage in development environments.
\add{In our current implementation, we utilize the \emph{Program Structure Interface} (PSI)~\cite{PSI}, a powerful toolkit for syntactic and semantic analysis, to gather compiler information for the \emph{Locator} and perform AST analysis for the \emph{Collector}.}

In addition, \emph{WIRL} employs an iterative strategy to perform context-aware code wiring. To balance latency, computational cost, and effectiveness, we empirically set the maximum number of iterations at two.

\subsection{RQ1: Outperform the SOTA}
\begin{table}[]
\centering
\caption{Comparison against Baselines}
\begin{tabular}{@{}cccccc@{}}
\toprule
Baselines             & \#EM &\#Rec & $EM_{P}$ & $EM_{R}$ & Latency \\ \midrule
IDEA             & 79 &  189    & 41.8\%    & 35.7\%  &  \textbf{3.7ms}\\ 
ExampleStack      & 4 &   10   &  40.0\%    &  1.8\%  & 729.1ms\\ 
CoEdPilot         & 21 &  32   &  65.6\%    &  9.5\% & 2,417.4ms\\ 
GrACE        & 163 & 184  &  88.6\%    &  73.8\%      & 2,873.4ms \\ \midrule
GPT-4o-mini       & 101    & 145      &69.7\% & 45.7\% & 3,930.4ms\\
Qwen2.5-Coder-14B      & 110 &  157       & 70.1\%           & 49.8\% & 4,624.5ms\\
Qwen2.5-Coder-32B      & 149 &  197        & 75.6\%           & 67.4\%  &6,799.4ms\\
Qwen-max       & 124  & 159          & 78.0\%    & 56.1\% &5,393.6ms \\
DeepSeek-V3      & 129 &  171       & 75.4\%    & 58.4\% &11,489.9ms\\ \midrule
\textbf{WIRL}        &   \textbf{199}  &  \textbf{217}   & \textbf{91.7\%}         & \textbf{90.0\%}   & 5,255.6ms    \\ \bottomrule
\end{tabular}
\label{tab:Performance}
\end{table}

In Table~\ref{tab:Performance}, the first column lists the evaluated approaches. The second and third columns indicate, respectively, the number of cases in which the recommended code elements exactly match the ground truth and the number of cases in which a recommendation is made by the approach. \add{The fourth and fifth columns}, $EM_{Precision}$ and $EM_{Recall}$, report how often the recommendations are correct and how many of the correct substitutions are successfully identified, respectively (see Section~\ref{subsec:Metrics} for detailed definitions).
From Table~\ref{tab:Performance}, we make the following observations:
\begin{itemize}[leftmargin=1em]
    \item \emph{WIRL} outperforms all selected baselines by a significant margin in both $EM_{Recall}$ and $EM_{Precision}$. It produces the highest number of recommendations (217) while also achieving the highest number of exact matches (199). The resulting $EM_{Recall}$ and $EM_{Precision}$ scores of 90.0\% and 91.7\%, respectively, reflect its strong and reliable performance in real-world development scenarios.
    \item Compared to \emph{IDEA}, \emph{WIRL} achieves improvements of 54.3 percentage points in $EM_{Recall}$ and 49.9 percentage points in $EM_{Precision}$. Against \emph{ExampleStack}, it shows even greater gains: 88.2 and 51.7 percentage points in $EM_{Recall}$ and $EM_{Precision}$, respectively. \add{Compared to advanced code editing approaches, i.e., \emph{GrACE} and \emph{CoEdPilot}, \emph{WIRL} achieves improvements of 16.2 and 80.5 percentage points in $EM_{Recall}$ and 3.1 and 26.1 percentage points in $EM_{Precision}$, respectively.} When benchmarked against the advanced LLM model \emph{Qwen2.5-Coder-32B}, \emph{WIRL} still delivers notable improvements of 22.6 and 13.7 percentage points, further demonstrating its superior effectiveness.
    \item \emph{WIRL} also substantially improves upon the performance of directly using a raw LLM (\emph{Qwen2.5-Coder-14B}). Without the academic design described in Section~\ref{sec:Approach}, the baseline LLM achieves only 49.8\% $EM_{Recall}$ and 70.1\% $EM_{Precision}$. With the integration of the external toolkit and structured prompt design, \emph{WIRL} improves $EM_{Recall}$ and $EM_{Precision}$ by 40.2 and 21.6 percentage points, highlighting the value of its architectural innovations.

\end{itemize}

\subsubsection{\add{Analysis Over IDEA}}
First, IDEA's suggestion mechanism, designed for renaming, lacks effective ranking mechanism for code wiring. Our analysis revealed that while IDEA's suggestion list contained the correct substitution in 84.1\% of cases (159 out of 189), it was often buried among a large number of irrelevant options, sometimes over 100, requiring significant developer effort to locate. Only 49.7\% candidates in the top position are correct answers. This increases cognitive load and slows down the adaptation process. In contrast, \emph{WIRL} provides a single, high-confidence recommendation for each unresolved element with 91.7\% precision, turning a tedious search into a simple validation step.
Second, IDEA failed to generate any recommendations in 14.5\% of cases. These failures typically occurred when the unresolved elements are involved with complex expressions rather than simple variable s, a scenario that falls outside the scope of its renaming functionality. \emph{WIRL}, by leveraging static code analysis, successfully handles these more complex cases, demonstrating its broader applicability and robustness.

\subsubsection{\add{Analysis Over ExampleStack}}
The core limitation of \emph{ExampleStack} is its dependency on a pre-collected database of historical code modifications. Its effectiveness is therefore constrained to scenarios that have been seen before, which significantly hampers its performance on novel problems common in real-world development.
In our evaluation, only 6 code wiring instances within $CWEvaluation$ had similar historical modifications in \emph{ExampleStack}'s database~\cite{ExampleStack}. Across all test cases, \emph{ExampleStack} attempted 10 recommendations, with only 4 being correct.
In contrast, \emph{WIRL} leverages the reasoning capabilities of LLMs to generate solutions dynamically and from scratch for each specific context. This design allows \emph{WIRL} to operate independently of historical data, giving it far greater generalization and effectiveness across diverse and unseen code wiring scenarios.

\subsubsection{\add{Analysis Over Code Editing Methods}}
\add{As presented in Table~\ref{tab:Performance}, \emph{WIRL} surpasses the performance of advanced code editing methods like \emph{GrACE} and \emph{CoEdPilot}. This superiority can be attributed to two primary factors:
First, these code editing methods are fundamentally designed to leverage a rich history of prior code edits. The code wiring scenario, however, involves pasting code from external sources, a context that inherently lacks such a historical record. This foundational mismatch significantly curtails the effectiveness of code editing methods.
Second, the core methodologies of these tools are not well-suited for code wiring.
\emph{CoEdPilot}'s key modules for analyzing prior and subsequent edits are ineffective in this context, as there is little relevant edit history to analyze. Consequently, its generation module, which is fine-tuned specifically for edit-history-rich tasks, is ill-equipped for the distinct challenge of code wiring.
\emph{GrACE}, utilizing GPT-4o in a zero-shot setting, achieves impressive results by combining the powerful completion capabilities of LLMs with a carefully engineered prompt. This design philosophy is highly analogous to that of \emph{WIRL}, and its success further validates the effectiveness of our approach.}

\subsubsection{\add{Analysis Over Raw LLMs}}
\begin{figure}[]
    \centering
\includegraphics[width=\linewidth,clip]{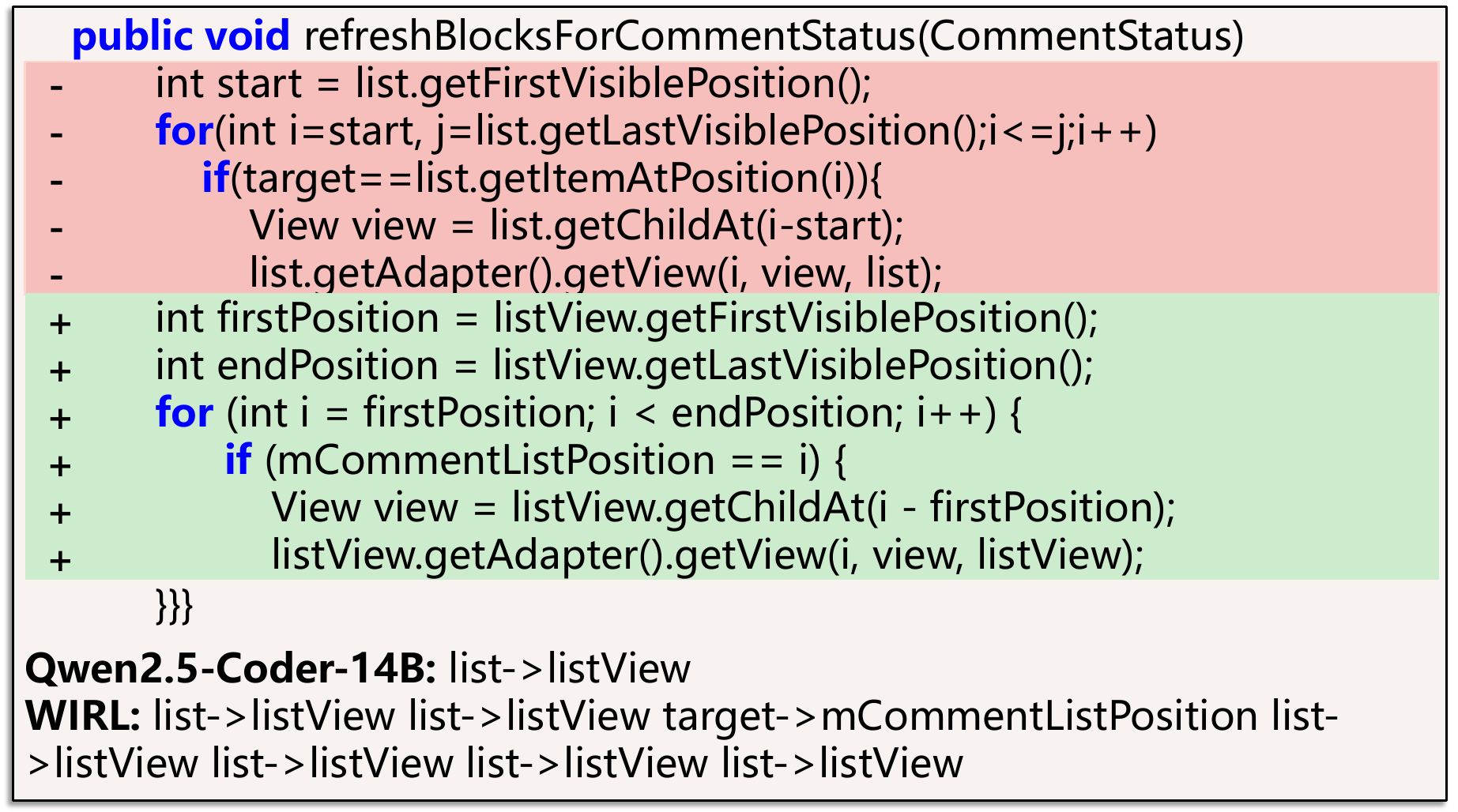}
    \caption{Example of Identification Failure for Raw LLM}
    \label{fig:Example1}
    \vspace{-1em}
\end{figure}

\emph{WIRL}'s superiority over raw LLMs is evident in both recall and precision.
The lower $EM_{Recall}$ of raw LLMs can be attributed to two main factors.
First, their fixed context windows prevent them from processing large classes that exceed token limits, resulting in complete failure.
Second, they often fail to identify all unresolved elements within the code. As illustrated in Fig.~\ref{fig:Example1}, \emph{Qwen2.5-Coder-14B} successfully identified \emph{``list''} and recommended \emph{``listView''}, but it failed to detect \emph{``target''}. In contrast, \emph{WIRL} uses compiler information to reliably identify all unresolved elements, overcoming these limitations.

\begin{figure}[]
    \centering
\includegraphics[width=\linewidth,clip]{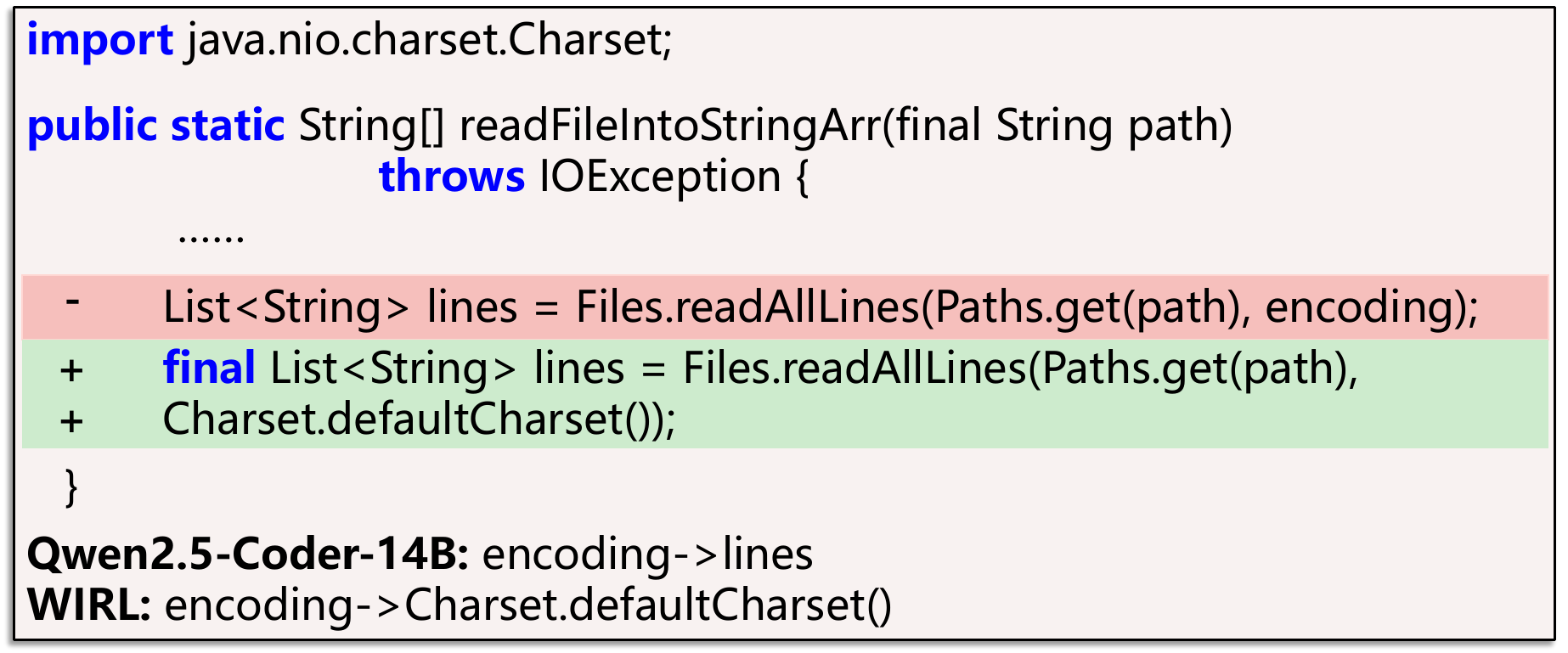}
    \caption{Example of Recommendation Failure for Raw LLM}
    \label{fig:Example2}
    \vspace{-1em}
\end{figure}

The reduced $EM_{Precision}$ of raw LLMs stems from their difficulty in analyzing complex contexts, especially when the substitution is not a simple variable but requires synthesizing code, e.g., a method call. As shown in Fig.~\ref{fig:Example2}, 
 a raw LLM is unlikely to deduce the correct substitution \emph{``Charset.defaultCharset()''} because it lacks awareness of the class's available static methods. In contrast, \emph{WIRL} addresses this limitation by leveraging its toolkit to first identify the required type (e.g., \emph{``Charset''}) and then retrieve relevant methods (e.g., \emph{``defaultCharset()''}), supplying the LLM with the necessary information to make a correct recommendation. Additionally, raw LLMs sometimes generate extraneous substitutions, increasing false positives and further lowering precision.

\subsubsection{\add{Failure Analysis for WIRL}}~\label{subsubsec:FailureAnalysis}
\begin{figure}[]
    \centering
\includegraphics[width=\linewidth,clip]{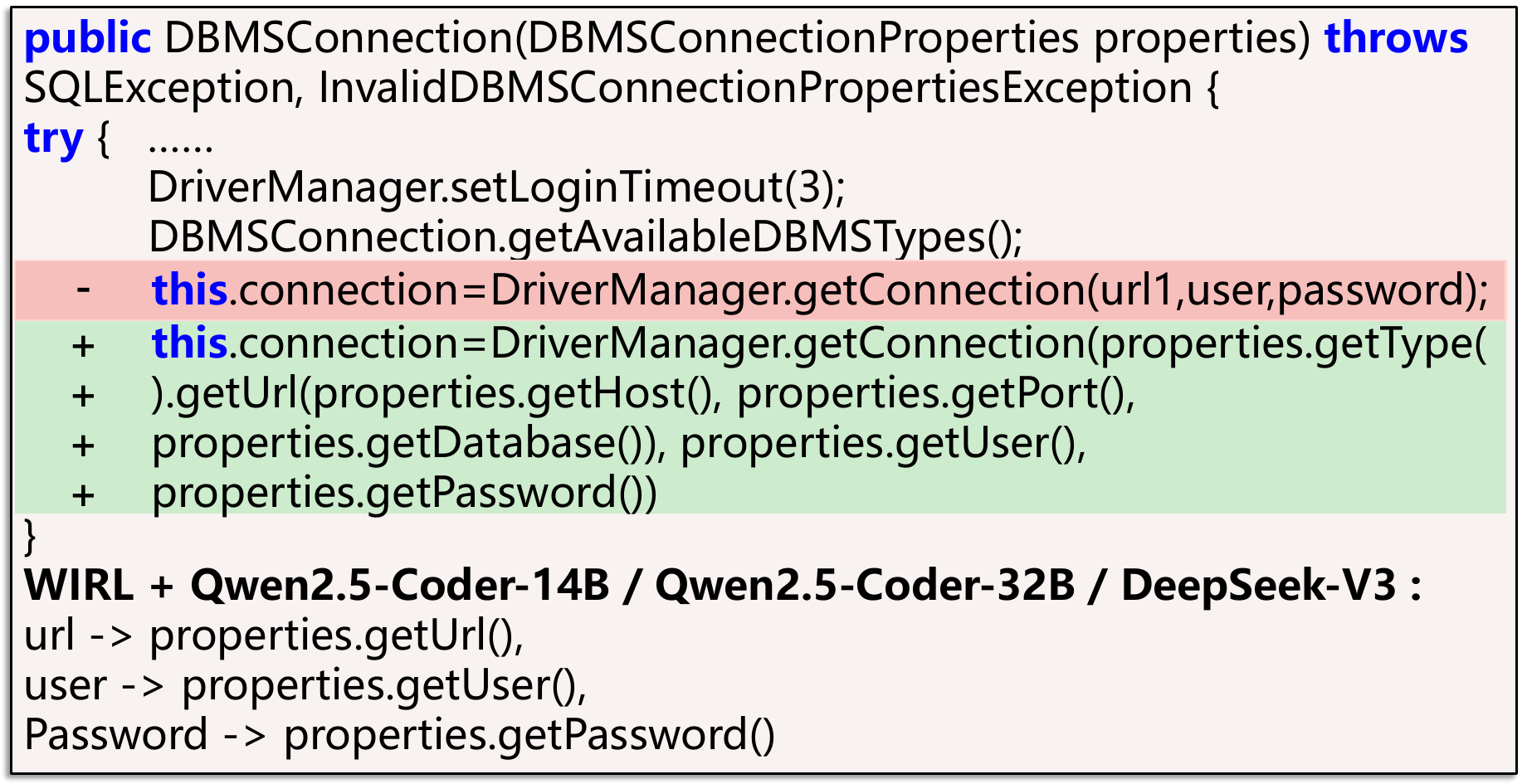}
    \caption{Example of Recommendation Failure for WIRL}
    \label{fig:Example3}
    \vspace{-1em}
\end{figure}

\add{While \emph{WIRL} demonstrates strong performance, analyzing its failures reveals its current limitations. We use a representative case testing with different base models (Fig.~\ref{fig:Example3}) to illustrate this boundary.
In this case, \emph{WIRL} must resolve three elements: \emph{``url1''}, \emph{``user''}, and \emph{``password''}. It successfully maps the simpler elements, \emph{``user''} and \emph{``password''}, to their corresponding context variables, \emph{``properties.getUser()''} and \emph{``properties.getPassword()''}. This highlights its proficiency in handling direct mappings.
The failure occurred with the element \emph{``url1''}. Its correct substitution is not a simple variable but a deeply nested chain of method invocations. Constructing this requires advanced, multi-step compositional reasoning, a known frontier challenge for current LLMs and agent systems. The model needs to synthesize a complex code structure rather than just perform a direct mapping.
This failure across various base models indicates that the limitation lies not only in the LLM's capabilities but also in the current agent-tool interaction paradigm, which is less effective for tasks requiring the synthesis of complex, multi-step code from scratch.}

\subsection{RQ2: Time Efficiency}

In this research question, we examine the time efficiency of \emph{WIRL} compared to other baseline approaches on the $CWEvaluation$. \add{The comparison results are summarized in Table~\ref{tab:Performance}, where the last column reports the average execution time per \emph{code wiring} instance.} It is worth noting that the reported time costs of raw LLMs are averaged over five independent runs to ensure robustness and consistency.
\add{From Table~\ref{tab:Performance}}, we make the following observations:
\begin{itemize}[leftmargin=1em]
    \item \emph{WIRL} required only marginally more time (approximately 0.6 seconds) to generate a recommendation compared to \emph{Qwen2.5-Coder-14B}, and it demonstrated greater time efficiency than all other evaluated LLMs except \emph{GPT-4o-mini}. This efficiency can be attributed to its use of a relatively compact LLM (14B parameters), whereas other models incur longer response times due to their significantly larger parameter sizes. For example, \emph{DeepSeek-V3}, with 671B parameters, exhibited the highest latency, averaging 11.5 seconds per recommendation.
    \item Although \emph{IDEA} and \emph{ExampleStack} achieved superior performance in terms of time cost, they fall short in ensuring recommendation accuracy. We argue that \emph{WIRL} offers a more practical trade-off between recommendation quality and acceptable latency, making it well-suited for real-world development workflows.
    \item \add{Code editing methods, such as \emph{GrACE} and \emph{CoEdPilot}, exhibit moderate time efficiency, placing them between the fast, heuristic-based methods and the slower, raw LLM baselines. This performance is attributable to their reliance on pretrained language models, which are more computationally intensive than simple heuristics.}
\end{itemize}

\begin{figure}[]
    \centering
\includegraphics[width=\linewidth,clip]{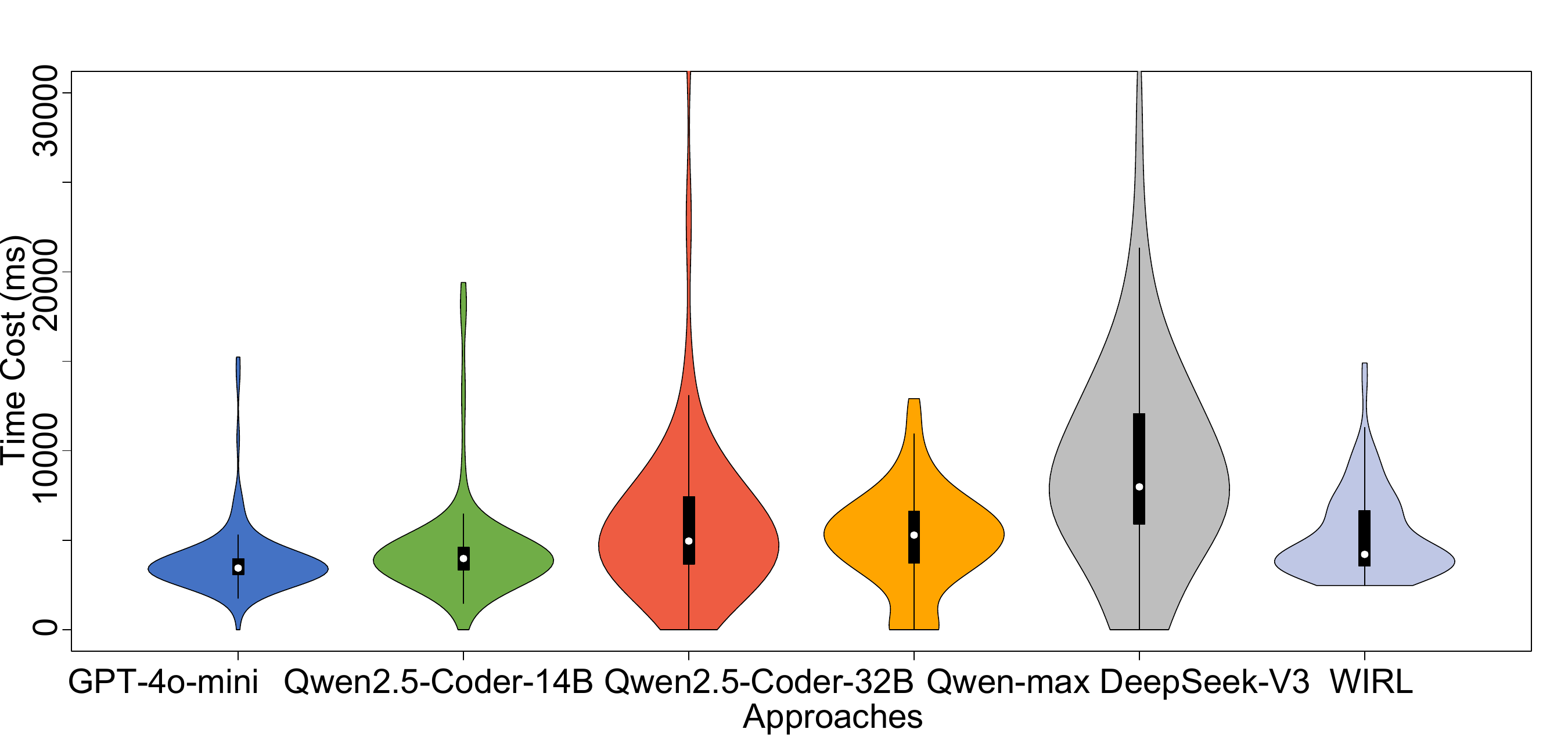}
    \caption{Time Cost Distribution}
    \label{fig:TimeCostViolin}
    \vspace{-1em}
\end{figure}

To gain deeper insights into the time cost distribution of \emph{WIRL} and the raw LLMs, we visualized the results using a violin plot, as shown in Fig.~\ref{fig:TimeCostViolin}. The plot reveals that \emph{WIRL} is capable of handling the majority of \emph{code wiring} instances in under 5 seconds. Its median time cost is 4,201.5ms, which is comparable to that of \emph{Qwen2.5-Coder-14B} (3,977.5ms), and substantially lower than that of \emph{DeepSeek-V3} (7,983.5ms). This demonstrates \emph{WIRL}’s ability to deliver high-quality recommendations with competitive latency.


\subsection{RQ3: Token Consumption and Monetary Cost}
\begin{table*}[]
\centering
\caption{Token Consumption, and Monetary Cost}
\begin{tabular}{@{}ccccc@{}}
\toprule
LLMs             & Avg. \#Input Tokens & Avg. \#Output Tokens & Avg. \#Total Tokens  & Avg. Monetary Cost (USD)  \\ \midrule
GPT-4o-mini      & 2,769.4              & 27.4 &  2,796.8  & 0.0004 \\
Qwen2.5-Coder-14B       & 2,769.4   &  28.3    &   2,797.7        & 0.0008
\\
Qwen2.5-Coder-32B     &  2,769.4  & 43.8       &   2,813.2        & 0.0008
\\
Qwen-max       &  2,769.4  &  24.9   &  2,794.3     & 0.0010       \\
DeepSeek-V3       &  2,769.4  &  22.0   &  2,791.4    &  0.0008      \\ \midrule
WIRL      & 3,006.7   &  108.5    & 3,115.2    & 0.0009                         \\ \bottomrule
\end{tabular}
\label{tab:Token}
\end{table*}
In this research question, we evaluate the token usage and monetary cost of \emph{WIRL} compared to other LLM-based baselines on the $CWEvaluation$. The comparison results are summarized in Table~\ref{tab:Token}. \add{Columns 2-4} report the average number of input tokens, output tokens, and total tokens per code wiring instance, respectively, while the last column shows the corresponding average monetary cost. All figures are averaged over five independent experimental runs.

As shown in Table~\ref{tab:Token}, \emph{WIRL} consumes slightly more tokens than \emph{Qwen2.5-Coder-14B}, with an average increase of 237.3 input tokens and 80.2 output tokens. This results in a marginally higher monetary cost, i.e., approximately 0.0001 USD more per instance. Surprisingly, although \emph{GPT-4o-mini} incurs the lowest monetary cost, its performance is relatively poor \add{(refer Table~\ref{tab:Performance})}. The increase in token usage is attributed to \emph{WIRL}'s iterative design, where it progressively collects and integrates contextual information into the prompt before generating a final recommendation. As the output from one iteration is passed as input to the next, the cumulative token count naturally exceeds that of the single-pass raw LLMs. 

\begin{figure}[]
    \centering
\includegraphics[width=\linewidth,clip]{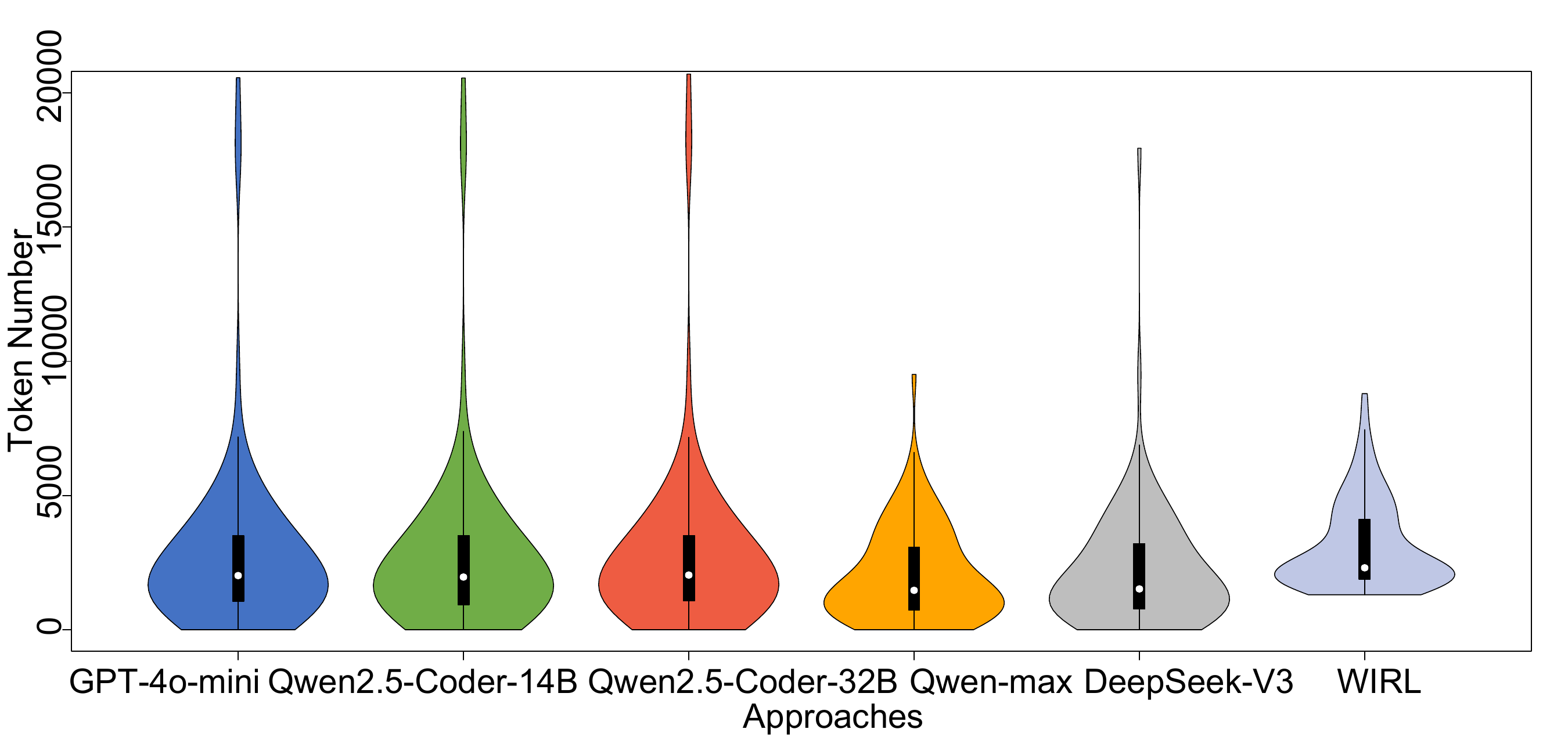}
    \caption{Token Number Distribution}
    \label{fig:TokenViolin}
    \vspace{-1em}
\end{figure}

To further examine the distribution of token consumption for \emph{WIRL} and the LLM baselines, we present a violin plot in Fig.~\ref{fig:TokenViolin}. The results indicate that \emph{WIRL} handles the majority of cases using fewer than 2,500 tokens, with a median token count of 2,306.5 per code wiring instance, which is comparable to that of \emph{Qwen2.5-Coder-14B} having a median of 1,962.5 tokens. Additionally, both \emph{WIRL} and \emph{Qwen-max} exhibit stable token usage, with minimal outlier points. Notably, \emph{Qwen-max} shows consistent performance in both time and token consumption, largely because it omits cases that exceed its maximum token limit.

\subsection{\add{RQ4: Ablation Study}}
\begin{table}[]
\centering
\caption{Ablation Study}
\begin{tabular}{@{}cccccc@{}}
\toprule
Variants     & \#EM &\#Rec & $EM_{P}$ & $EM_{R}$ & Latency \\ \midrule
\textbf{WIRL}  &\textbf{199}  &  \textbf{217}   & \textbf{91.7\%}  & \textbf{90.0\%}  & 5,255.6 ms    \\ 
w/o Locator  &   110  &  157   & 70.1\%         & 49.8\% & \textbf{4,624.5 ms} \\
w/o Collector  &   169  &  190   & 88.9\%         & 76.5\% & 5,048.2 ms\\
w/o Completer  &   146  &  189   & 77.2\%         & 66.1\% & 4,827.1 ms\\ 
w/o State Machine &   198  &  217  &   91.2\%   &89.6\%  & 7,297.1 ms\\
w/o Filtering Tools &   179  &  205 & 87.3\%    & 81.0\% & 5,679.7 ms\\
w/o Retrieval Tools &   189  &  216    &  87.5\%     &85.5\%  & 5,820.3 ms\\ 
 \bottomrule
\end{tabular}
\label{tab:AblationStudy}
\end{table}
\add{To investigate the impact of each design choice on overall performance, we conducted an ablation study by creating several variants of \emph{WIRL}. We first individually removed the three types of customized tools, \emph{Locator}, \emph{Collector}, and \emph{Completer}, to assess their respective contributions (Rows 3-5 in Table~\ref{tab:AblationStudy}). Next, we evaluated the impact of key architectural components. We replaced the state machine with a fully agent-guided exploration strategy and individually removed the filtering tools (i.e., reserve\_type\_compatible\_ones and sort\_by\_literal\_similarity), and the retrieval tools (retrieve\_identical\_function\_call and get\_method\_names). 
From Table~\ref{tab:AblationStudy}, we make the following observations:
\begin{itemize}[leftmargin=1em]
    \item  \textbf{Tool Contributions:} Three tool types contribute differently to \emph{WIRL}'s performance. The \emph{locator} tools are the most critical, followed by the \emph{completer} and then the \emph{collector} tools. The paramount importance of the \emph{locator} tools is because both the \emph{collector} and \emph{completer} depend on the precise location information they provide. Without \emph{locator}, the system effectively degrades to a raw LLM.
    \item \textbf{Architectural Necessity:} All architectural components are integral to the overall performance. The data clearly shows that removing any component, the state machine, filtering tools, or retrieval tools, results in a performance degradation. This highlights their crucial roles in providing valid contextual information to the agent. While the time efficiency remains stable across most variants, the removal of the state machine presents a notable exception.
    \item \textbf{State Machine Efficiency:} The moderate latency increase observed when removing the state machine is a direct result of a proactive design choice to prevent aimless exploration, a common pitfall in agentic systems. We constrained the maximum iteration number to two, which limits the observable latency penalty in this study. Without this constraint, removing the state machine would lead to prolonged, multi-step exploration cycles and a significantly more substantial, and likely unacceptable, increase in latency.
\end{itemize}
The ablation study results confirm that each component is integral to \emph{WIRL}'s performance and the architecture designs are also the key contributors to the efficiency of \emph{WIRL}.}

\section{Discussion}\label{sec:Discussion}
\subsection{\add{Placing WIRL in the Broader Context of Code Adaptation}}
\add{Despite its specific focus, code wiring represents a frequent, critical, and foundational aspect of code adaptation. This claim is supported by two key factors:
\begin{itemize}[leftmargin=1em]
    \item \textbf{Prevalence in Adaptations:} The empirical study by Zhang et al.~\cite{Zhang2024How} on real-world code adaptations provides strong evidence for this. Their study revealed that a remarkable 88.0\% of code reuses require adaptation. Through our further analysis, over one quarter (27.3\%) of those adaptations involve code wiring operation,  underscoring its high prevalence. Additionally, the study highlights that variable references and method invocations are central to the adaptation process, with 77.2\% of statement-level code adaptations involving method invocations or variable references.
    \item \textbf{Task Complexity:} While direct variable-to-variable mapping is relatively straightforward, the core challenge lies in mapping a variable to a complex expression, such as a deeply nested method invocation. This elevates the task from simple pattern matching to one requiring deeper program semantic understanding. A complex example is presented in Fig.~\ref{fig:Example3} and the analysis is presented in Section~\ref{subsubsec:FailureAnalysis}.
\end{itemize}}
 
\subsection{\add{Applicability Analysis}}
\add{We evaluate WIRL's real-world applicability across three key dimensions:
\begin{itemize}[leftmargin=1em]
\item \textbf{Accuracy:} With a precision of 91.7\%, WIRL significantly reduces the developer's cognitive load by turning a tedious and error-prone manual adaptation process into a simple inspection and approval task in nine out of ten cases. This accuracy substantially outperforms the 71.8\% of the previous state-of-the-art approach~\cite{zhang2024instructinteractexploringeliciting} in code snippet adaptation.
\item \textbf{Latency:} WIRL introduces a minimal computational overhead of only 0.6 seconds on top of the base model's inference time. The average 5.25-second recommendation time, inflated by our experimental setup using online API calls, would be significantly reduced in a production environment with a locally deployed model or a dedicated API, making it well-suited for practical development workflows. To support self-hosted models or private APIs, the implementation of \emph{WIRL}~\cite{WIRLWithLicense} allows for the configuration of a custom endpoint URL via the IDEA settings panel at File - Settings - Tools - WIRL AI Settings.
\item \textbf{Cost:} Economically, WIRL is highly efficient, with an average cost of just \$0.0009 per recommendation based on public API pricing. This negligible cost yields a high return on investment by saving valuable developer time.
\end{itemize}}

\subsection{\add{Dataset Representativeness Analysis}}
\add{Although our dataset, $CWEvaluation$, is modest in size, its quality and representativeness are justified by:
\begin{itemize}[leftmargin=1em]
\item \textbf{Ecological Validity:} $CWEvaluation$ is derived from genuine code reuse instances from the SOTorrent database filtered and validated by Zhang et al.~\cite{Zhang2024How}, ensuring it reflects authentic development practices.
\item \textbf{Project and Domain Diversity:} It includes 29 distinct open-source projects from a wide range of domains, e.g., foundational libraries and popular mobile applications, enhancing the generalizability of our findings.
\item \textbf{Structural Diversity:} The code snippets' structural properties, e.g., LOC, are comparable to those in larger datasets~\cite{Zhang2024How}, and a distributional analysis can be found in our online website~\cite{LOCDistributionalAnalysis}. Besides, $CWEvaluation$ covers both file-level and project-level adaptation scenarios, ensuring its diversity and complexity.
\end{itemize}}

\subsection{\add{Versatility Analysis}}
\add{\emph{WIRL} is designed to be language-agnostic. Its core LLM-based reasoning and agentic control flow are language-independent. To adapt WIRL to a new language, only two language-specific components in its toolkit require modification: the \emph{Locator}, which uses compiler information to find unresolved variables, and the \emph{Collector}, which leverages Abstract Syntax Tree (AST) analysis to gather contextual information. This modular design ensures high generalizability.}
\subsection{Threats to Validity}
The primary threat to \textbf{external validity} is the dataset's size. We mitigate this by using a public, manually inspected dataset~\cite{Zhang2024How} and by open-sourcing our data and results~\cite{WIRL} to encourage replication. A threat to \textbf{construct validity} is potential bias in identifying \emph{code wiring} instances. To address this, two authors independently inspected the data, achieving a high inter-rater agreement (Cohen’s kappa of 0.81), with all discrepancies resolved through discussion.
\subsection{Limitations}
\add{\emph{WIRL} has three main limitations. First, in the current implementation, if WIRL's analysis does not identify any suitable existing variables with high confidence, it will not make a replacement recommendation. Instead, it will inform the developer that no suitable match was found, implicitly guiding them to declare a new variable. For future work, we plan to enhance WIRL to not only detect this scenario but also suggest contextually appropriate names and types for the new variable declaration. Second, our dataset, $CWEvaluation$, lacks more complex wiring scenarios, such as those with semantic ambiguities or many-to-many mappings, due to the laborious nature of manual collection. We plan to develop a semi-automated approach to expand the dataset in the future.} Finally, \emph{WIRL} is currently implemented and evaluated only for Java, and extending it to other languages remains future work.

\section{Related Work}\label{sec:RelatedWork}
\subsection{Code Snippet Adaptation} 
Research into code snippet adaptation began with empirical studies classifying the reuse and modification patterns developers employ when integrating external code~\cite{Wu2019How, Chen2024How, Zhang2024How}. Early automated tools like \emph{CSNIPPEX}\cite{Terragni2016CSNIPPEX} and \emph{NLP2Testable}\cite{Reid2020Optimising} addressed unresolved variables by declaring them with default values, a solution insufficient for complex, context-dependent adaptations. A more significant body of work leverages code clones and existing examples. Tools such as \emph{SniptMatch}\cite{Wightman2012SnipMatch}, \emph{ExampleStack}\cite{Zhang2019Analyze}, \emph{CCDemon}\cite{Lin2015Clone}, \emph{MICoDe}\cite{Lin2017Mining}, and \emph{EUKLAS}\cite{Dorner2014EUKLAS} recommend adaptations by matching snippets to local or online code repositories. However, their reliance on finding similar, previously adapted code limits their applicability to novel scenarios. Another strategy frames adaptation as a de-obfuscation task, masking and recovering identifiers using context-aware models\cite{allamanis2017smartpastelearningadaptsource,Liu2023AdaptivePaste}. Other tools address related but distinct goals, such as assessing code compatibility~\cite{Cottrell2008JigSaw} or converting snippets into methods~\cite{Terragni2021APIzation}. 

In contrast, \emph{WIRL} focuses on resolving unresolved variables in new contexts without depending on prior examples.

\subsection{\add{Code Editing} }
\add{Recent advancements in code editing also inform our approach. Some models are pre-trained specifically for code modification, such as CoditT5~\cite{CoditT5}, which models edits directly, and CCT5~\cite{CCT5}, which learns from code change datasets. Other methods leverage edit history and structural context: C3PO~\cite{C3PO} represents edits as paths in the Abstract Syntax Tree, while Overwatch~\cite{OverWatch} learns from sequences of developer actions. More advanced LLM-based approaches like GrACE~\cite{GrACE} and CoEdPilot~\cite{CoEdPilot} condition edit generation on prior associated edits to better capture developer intent. }

\add{While powerful, these methods differ from \emph{WIRL} in two key ways. First, the edit location of code wiring problem is often a deterministic problem guided by compiler errors, whereas general code editing involves less predictable changes across multiple files. Second, advanced code editing approaches are dependent on a rich history of prior edits, information that is inherently scarce in the context of code wiring.}

\subsection{LLM-based Agents for Software Engineering} 
LLM-based agents are increasingly used to automate complex software engineering tasks through iterative reasoning and tool use~\cite{jin2024llmsllmbasedagentssoftware, wang2024survey,Zhang2025Deep}. They have been successfully applied to code generation and testing~\cite{huang2024agentcodermultiagentbasedcodegeneration, lin2024soen101codegenerationemulating, Zhang2024AutoCodeRover, Yang2024SWEAgent}, code maintenance and repair~\cite{batole2025localizeAgent, Bouzenia2024RepairAgent, Ke2025NIODebugger}, and automated issue resolution~\cite{Tao2024MAGIS}. 

Despite these advancements, no prior work addresses context-aware code wiring for pasted snippets. \emph{WIRL} fills this gap as the first framework to integrate agent-based planning and a customized toolkit for automated variable resolution.

\section{Conclusions and Future Work}\label{sec:Conclusion}
We introduce \emph{WIRL}, a novel LLM-based agent designed to facilitate context-aware code wiring, an essential but underexplored aspect of the copy-paste-modify paradigm in software development. Unlike prior approaches 
, \emph{WIRL} keeps a dynamically updated prompt to collect and record useful context information through a hybrid architecture that combines structured tool invocation with autonomous reasoning capabilities of LLMs. By formulating the adaptation task as an RAG infilling problem, \emph{WIRL} aligns closely with the natural strengths of LLMs in code completion.
Our evaluation demonstrates that \emph{WIRL} significantly outperforms commercial IDEs, \add{advanced code editing approaches,} and advanced LLMs. 

Looking ahead, we envision \emph{WIRL} as a stepping stone toward more general-purpose adaptation agents that can support a broader range of integration and transformation tasks. 

\section{Data Availability}
The replication package, including tools and data, is publicly available in~\cite{WIRL}.

\section*{Acknowledgment}{The authors thank the reviewers  for their insightful comments and constructive suggestions. This work was partially supported by the National Natural Science Foundation of China (62232003 and 62172037).}

\newpage

\bibliographystyle{IEEEtran}
\balance
\bibliography{reference}
\end{document}